\documentclass[times,5p]{elsarticle}
\usepackage{graphicx}
\usepackage{amssymb}
\usepackage{mathrsfs}
\usepackage{amsmath}
\usepackage{comment}
\usepackage{amsbsy}
\def\dbar{{\mathchar'26\mkern-12mu d}}
\journal{Physics Letter B}
\begin{document}
\begin{frontmatter}
\title{The role of photon  polarization modes in the magnetization and instability  of the vacuum in a supercritical field}
\author{Selym Villalba-Ch\'avez}
\address{Institute of Physics, University of Graz, Universit\"atsplatz 5, 8010, Graz, Austria.}
\begin{abstract}
The response of the QED vacuum  in an asymptotically large electromagnetic field is studied. In this regime the vacuum energy is strongly influenced  by the vacuum polarization effect. The  possible interaction between  the virtual electromagnetic radiation and  a superstrong  magnetic field  suggests  that a background of virtual  photons  is a source of magnetization to the whole vacuum. The corresponding contribution to the vacuum magnetization density is  determined by considering the individual contribution of each vacuum polarization eigenmode in the Euler-Heisenberg Lagrangian.  Additional issues concerning  the transverse pressures are analyzed. We also study the case in which the vacuum is occupied by a superstrong electric field. It is discussed  that, in addition to the electron-positron pairs, the vacuum could create  photons with  different propagation modes.  The possible relation between the emission of photons and the birefringent character of the vacuum is  shown as well.
\end{abstract}
\end{frontmatter}

\section{Introduction}

Whilst there is some  evidence  that  very large magnetic fields
$\vert\textbf{B}\vert\gg \rm B_c,$ $\rm B_c=m^2/e=4.42\cdot 10^{13}
G$\footnote{Hereafter $\rm m$ and $\rm e$ are the electron mass  and
charge, respectively.} exist in  stellar objects identified as
neutron stars  \cite{Manchester,Kouveliotou,Bloom},  its  origin and
evolution  remains  poorly understood  \cite{Uzdensky:2009cg}.
Some investigations in this area  provide theoretical  evidence that
$\vert\textbf{B}\vert$ might be generated due to gravitational and
rotational effects, whereas other theories estimate that is
self-consistent due to the  Bose-Einstein condensation  of charged
and  neutral boson gases in a superstrong magnetic field
\cite{Chaichian:1999gd,Martinez:2003dz,PerezRojas:2004ip,PerezRojas:2004in}.
In this framework  the nonlinear QED-vacuum  possesses the
properties of a  paramagnetic medium  and constitutes  a source of
magnetization, induced by the external magnetic field. Its
properties  are primarily  determined by the vacuum energy of
virtual electron-positron pairs. Because of this,  a negative
pressure transversal to the external field is generated
\cite{hugo4} in similarity  with the Casimir effect between metallic
plates \cite{bordag}.  Moreover,  the vacuum occupied by the
external field turns out to be an ``exotic''  scenario in which
processes  like  photon splitting \cite{adler,adler1}  and photon
capture \cite{shabad3,shabad3v,Herold:1985zz} could take place.
These two  phenomena  depend  on the photon dispersion relation
which differs  from the light cone, due to  vacuum polarization
effects \cite{shabad1,shabad2,shabad5,shabad6}.  As a result,  the
issue of  light propagation in empty space, in the presence of an
external magnetic field, is  similar to the dispersion of light in
an anisotropic ``medium''.

\begin{figure}[tbp]
\begin{center}
\includegraphics[width=2.5in]{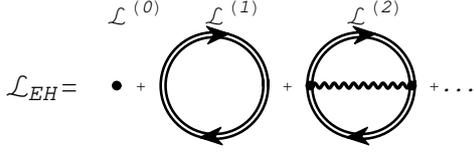}
\caption{\label{fig:mb000} Two-loop expansion  of the
Euler-Heisenberg Lagrangian. The double lines represent the
electron-positron Green's functions, whereas the wavy line refers to
the photon. Here $\mathcal{L}^{(0)}$ is the free  Maxwell
Lagrangian, $\mathcal{L}^{(1)}$ represents  the one-loop which gives
the contribution of  the virtual free electron-positron pairs
created and annihilated spontaneously in vacuum and interacting with
the external field. The radiative corrections (involved in
$\mathcal{L}^{(2)}$) emerge from two-loop due to exchange of the
virtual photons.}
\end{center}
\end{figure}

The  phenomenological aspects associated with  this problem    have
been studied for a long time. In the meanwhile,  other  features of
nonlinear electrodynamics in a superstrong magnetic field  have been
studied   such as  the  dimensional reduction of the Coulomb
potential \cite{shabad7,shabad8,shabadtalk,Sadooghi} and the
possible existence of a  photon anomalous magnetic moment
\cite{selym1}.  However, due to  the vacuum polarization  effect,
virtual photons can carry  a magnetization as well. As a
consequence, they might be a source of magnetism  to the whole
vacuum.  Motivated by this idea, we  address the question in which
way the virtual electromagnetic radiation   contributes to  the
vacuum magnetization and therefore to increase the external field
strength.  The magnetic  properties of the vacuum have been studied
in  \cite{hugo4,hugo3,hugo5,hugo6} for weak ($\vert\bf B\vert\ll \rm
B_c$) and  moderate fields  ($\vert\bf B\vert\sim \rm  B_c$) in
one-loop approximation of the Euler-Heisenberg Lagrangian
\cite{euler} which   involves  the contribution from  virtual
electron-positron pairs.   The  contribution of virtual photons,
created and annihilated spontaneously in the vacuum and  interacting
with $\textbf{B}$ by means of  $\Pi_{\mu\nu}$, is contained within
the two-loop term of the Euler-Heisenberg Lagrangian (see Fig.
\ref{fig:mb000}). The latter  was  computed many years ago  by
Ritus \cite{ritus,ritus1} and   has been recalculated by several
authors as well  \cite{dittrich1,schubert,kors,gies}. In all these
works, however, it is really cumbersome to discern the individual
contributions given by each virtual photon propagation mode to the
Euler-Heisenberg Lagrangian which should  allow to determine the
magnetism and  pressure  associated with each form of virtual  mode.
In this Letter  we analyze these contributions separately for very
large magnetic fields  $(\vert\bf B\vert\gg\rm  B_c)$ since these
allow  to establish  relations between the  birefringence  of the
vacuum \cite{shabad6,Dittrich:1998gt,Heinzl:2006pn} and the global
properties of it.

Besides the strongly magnetized vacuum, there is   another
interesting  external field configuration which deserves  to be
analyzed: a  supercritical electric field $\vert\mathbf{E}\vert\gg
\mathrm{E}_c$  with  $\rm
E_c=\mathrm{m}^2/\mathrm{e}=1.3\cdot10^{16}\mathrm{V}/\mathrm{cm}$.
In this asymptotic region   the  Euler-Heisenberg Lagrangian
acquires  an imaginary term which   characterizes  the instability
of the  vacuum. This phenomenon  is  closely  related to  the
production of observable particles from the  own vacuum.  Certainly,
the creation of electron-positron pairs -the so-called Schwinger
mechanism-  turns out to be the most remarkable effect predicted
through this procedure \cite{euler,sauter,Schwinger}. However, the
imaginary part of this effective   Lagrangian is just a measure of
the vacuum decay and does neither give the  actual rate of
production of particles nor  the accessible decay channels
\cite{Cohen:2008wz}. Thereby not only the creation of
electron-positron pairs is  a plausible effect but also the emission
of observable photons \cite{ritus,ritus1}. The latter phenomenon was
analyzed  by   Gitman, Fradkin  and Shvartsman
\cite{Gitman:1986xr,gitman1,gitman2}. Their results showed that the
total probability of photon emission from the vacuum, accompanied by
the creation of an arbitrary number of electron-positron pairs, is
connected to the decay probability of the vacuum and thus to the
imaginary part arising from the  two-loop term of the
Euler-Heisenberg Lagrangian. In this context  the corresponding
decomposition  in terms of the vacuum polarization modes is
particularly illuminating because it  reveals that only two  of them
contribute to the vacuum instability.  It seems, therefore, that the
vacuum could create  photons with  different propagation modes, an
effect closely related to its own  birefringence.

\section{Preliminary remarks}

In a magnetized vacuum the spatial  symmetry is explicitly broken
by  the external field $\bf B.$ In this context, there is a
vectorial basis  $\flat^{(i)}_\mu$
\cite{shabad7,batalin,Shabad:2009ky}  which characterizes the vacuum
symmetry properties and    fulfills  both the orthogonality
condition: $
\flat_{\sigma}^{(i)}\flat^{\sigma(j)}=\delta^{ij}\left(\flat^{(i)}\right)^2$
 and  the completeness relation:
$\delta^{\mu}_{\ \ \nu}-\frac{k^\mu
k_\nu}{k^2}=\sum_{i=1}^3\flat^{\mu(i)}\flat_{\nu}^{(i)}/\left(\flat^{(i)}\right)^2.$
Explicitly, the basis vectors read  $\flat^{(1)}_\mu= k^2
\mathscr{F}^2_{\mu \lambda}k^\lambda-k_\mu (k\mathscr{F}^2 k)$,
$\flat^{(2)}_\mu=\tilde{\mathscr{F}}_{\mu \lambda}k^\lambda$,
$\flat^{(3)}_\mu=\mathscr{F}_{\mu \lambda}k^\lambda$ and
$\flat^{(4)}_\mu=k_\mu.$ These expressions  involve the external
field tensor  $\mathscr{F}_{\mu\nu}$ and its dual
$\tilde{\mathscr{F}}^{\mu\nu}=1/2\epsilon^{\mu\nu\rho\sigma}\mathscr{F}_{\rho\sigma}.$
In this basis, the vacuum polarization tensor is diagonal
\emph{i.e.}
\begin{equation}
\Pi_{\mu\nu}=\sum_{i=0}^4\varkappa_i\frac{\flat_\mu^{(i)}\rm
\flat_{\nu}^{(i)}}{\left(\flat^{(i)}\right)^2}
\end{equation}whereas  the dressed photon Green function can be expressed  as
\begin{equation}
\mathscr{D}_{\mu\nu}=\sum_{i=1}^3\frac{1}{\textrm{k}^2-\varkappa_i}\frac{\flat_\mu^{(i)}\flat_{\nu}^{(i)}}{\left(\flat^{(i)}\right)^2}+\frac{\zeta}{\rm k^2}\frac{\rm
k_\mu k_\nu}{\rm k^2}.
\label{photonpro}
\end{equation} Here the $\varkappa_i$  represent the $\Pi_{\mu\nu}-$eigenvalues
and $\zeta$ is the gauge parameter.   This diagonal decomposition
of  $\Pi_{\mu\nu}$ defines the energy spectrum of the
electromagnetic field which differs from the  isotropic vacuum ($\bf
B=0$).

Owing  to the transversality  property $(\rm k^\mu\Pi_{\mu\nu}=0),$
the  eigenvalue corresponding to the fourth eigenvector  vanishes
identically $(\varkappa^{(4)}=0).$ Furthermore, not all the
remaining eigenmodes are physical. In general, this depend on the
direction of wave propagation. To show this we  consider
$\flat^{(i)}_\mu( k)$ as the electromagnetic four vector describing
a photon. The corresponding electric and magnetic fields of each
mode are $\boldsymbol{ e}^{(i)}=i(\boldsymbol{k} \flat^{(i)}_0
-\omega^{(i)} \boldsymbol{\flat}^{(i)})$, $\boldsymbol{
b}^{(i)}=-i\boldsymbol{k}\times\boldsymbol{\flat}^{(i)}$. It follows
that the mode $i=3$ is a wave polarized in the transverse plane to
$\boldsymbol{k}$ whose electric
$\boldsymbol{e}^{(3)}\sim\boldsymbol{k}_\perp\times
\boldsymbol{n}_{\parallel}$ and magnetic
$\boldsymbol{b}^{(3)}\sim\boldsymbol{n}_{\parallel}
k_\perp^2-\boldsymbol{k}_\perp k_{\parallel}$ fields are orthogonal
to $\textbf{B}$ as well. Here the vectors $\boldsymbol{k}_{\perp}$
and $\boldsymbol{k}_{\parallel}$ are the components of
$\boldsymbol{k}$ across and along $\bf B$  with
$\boldsymbol{n}_\parallel= \mathbf{B}/\vert \mathbf{B}\vert$. For a
pure longitudinal propagation to the external field $k_\perp=0$, the
mode $\flat^{(2)}_\mu$ is a longitudinal and non-physical electric
wave $\boldsymbol{e}^{(2)}\sim\boldsymbol{n}_{\parallel}.$ On the
other hand,  $\flat^{(1)}_\mu$  is transverse since the associated
electric field is $\boldsymbol{e}^{(1)}\sim\boldsymbol{k}_\perp$
whereas the magnetic one
$\boldsymbol{b}^{(1)}\sim\boldsymbol{k}_\perp\times\boldsymbol{k}_\parallel$.
As a consequence, both   $\flat_{\mu}^{(1)}$ and $\flat_{\mu}^{(3)}$
represent  physical waves which may be combined to form a
circularly polarized transversal wave. In this case both modes
propagate along $\bf B$  with a  dispersion law  independents of the
magnetic field strength \cite{shabad2,shabad6,Shabad:2009ky}.

Now, if the photon propagation   involves a nonvanishing
trans\-versal momentum component $k_\perp\neq0,$ we are allowed to
perform the analysis in a Lorentz frame that  does not change the
value $ k_\perp$,  but gives $ k_\parallel=0$ and does not introduce
an external electric field. In this Lorentz frame, the   first
eigenmode $\flat_\mu^{(1)}$  becomes purely electric longitudinal
and a  non physical mode whereas $\flat_\mu^{(2)}$ is transverse.
Hence, for  a photon whose three-momentum  is directed at any
nonzero angle with the external magnetic field, the  two orthogonal
polarization states $\flat_\mu^{(2)}$ and $\flat_\mu^{(3)}$
propagate.  In this framework  the analytical structures of the
corresponding eigenvalues $\varkappa_{2,3}$ are different. As a
matter of fact, the vacuum behaves like  a birefringent medium with
refraction indices \cite{shabad6,Shabad:2009ky}
\begin{equation}
\eta_2=\frac{\vert\boldsymbol{k}\vert}{\omega_2}=\left(1+\frac{\varkappa_2}{\omega_2^2}\right)^{1/2}\ \ \mathrm{and}\ \ \eta_3=\frac{\vert\boldsymbol{k}\vert}{\omega_3}=\left(1+\frac{\varkappa_3}{\omega_3^2}\right)^{1/2}.
\end{equation} Here  $\omega_{2,3}$ are the corresponding
solution of the dispersion equations $k^2=\varkappa_{2,3}$ arising
from the poles of $\mathscr{D}_{\mu\nu}$.

Considering these aspects,  we  analyze the  Euler-Heisen\-berg  Lagrangian
\begin{equation}
\mathcal{L}_{\mathrm{EH}}=\mathcal{L}_{\mathrm{R}}^{(0)}+\mathcal{L}_{\mathrm{R}}^{(1)}+\ldots\label{therdp}
\end{equation} where $
\mathcal{L}_{\mathrm{R}}^{(0)}=-\frac{1}{2}\mathrm{B}^2$ is the free
renormalized Maxwell Lagrangian, whereas
$\mathcal{L}_{\mathrm{R}}^{(1)}$ denotes the  one loop  regularized
contribution  of virtual electron-positron pairs created and
annihilated spontaneously in vacuum and interacting with $\bf B$
\cite{euler}. In asymptotically large magnetic fields it reads
\begin{equation}
\mathcal{L}_{\mathrm{R}}^{(1)}(\mathfrak{b})\approx\frac{\rm m^4 \mathfrak{b}^2 }{24 \pi^2 \rm
}\left\{\ln\left(\frac{\mathfrak{b}}{\gamma \pi}\right)+\frac{6}{\pi^2}\zeta^{\prime}(2)\right\}.\label{1looprenorapp}
\end{equation} Here $\mathfrak{b}=\vert\mathbf{B}\vert/\mathrm{B}_c$,  $\ln(\gamma)=0.577\ldots$ denotes  the Euler constant whereas  $6\pi^{-2}\zeta^{\prime}(2)=-0.5699610\ldots$ and  $\zeta(x)$ is  the Riemann zeta-function.

\begin{figure}[!tp]
\begin{center}
\includegraphics[width=2.5in]{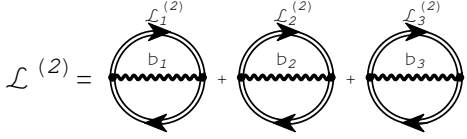}
\caption{\label{fig:mb0fgh00} Diagrammatic decomposition of
$\mathcal{L}^{(2)}$  in terms of the vacuum polarization
eigenmodes.}
\end{center}
\end{figure}

The  contribution  of a virtual photon interacting with external
field  by means of the vacuum polarization tensor is expressed as \cite{ritus,dittrich1,gies}
\begin{equation}
\mathcal{L}^{(2)}=\frac{i}{2}\int\frac{\rm
d^4k}{(2\pi)^4}\Pi_{\mu\nu}(\rm k)\mathscr{D}_0^{\mu\nu}(\rm k),\label{twoloopunrenor}
\end{equation} where $\mathscr{D}_0^{\mu\nu}(\rm k)$  denotes the  bare photon Green function which  is  obtained  by  neglecting  $\varkappa_i$ in the denominator of $\mathscr{D}_{\mu\nu}.$  Using this expression we find
\begin{equation}
\mathcal{L}^{(2)}=\frac{i}{2}\sum_{i=1}^3\int \dbar k\frac{\varkappa_i}{k^2},\label{preliminareuaqtax}
\end{equation}where $\dbar k\equiv d^4k/(2\pi)^4$. To obtain the above expression we have inserted the diagonal decomposition of $\Pi_{\mu\nu}$   into Eq. (\ref{twoloopunrenor}) and used the orthogonality condition.  Certainly, the quantity $\varkappa_i  k^{-2}$ represents the interaction energy density of  a  virtual photon propagation mode  with $\bf B$. Because  the analytical properties of the  $\varkappa_i$ differ
from each other the contribution of  each mode  will be different. Therefore,  the original two-loop graph can be  decomposed into three diagrams (see Fig.
\ref{fig:mb0fgh00})
whose  properties  are connected to the  birefringence property of the vacuum. Therefore,  each individual term  $\mathcal{L}_i^{(2)}=\frac{i}{2}\int \dbar k \varkappa_i/ k^2$ will be  studied separately.
As we  shall see very shortly, the  decomposition of $\mathcal{L}^{(2)}$ is also  valid in an external electric field in which  case  some evidence on the  possible  decay of the vacuum  into observable photons emerges.

Obviously,  the renormalized two-loop term decomposes as well: $\mathcal{L}_{\mathrm{R}}^{(2)}=\sum_{i=1}^3\mathcal{L}_{i\mathrm{R}}^{(2)}$  and the leading  behavior of $\mathcal{L}_{i\mathrm{R}}^{(2)}$ in an  asymptotically large magnetic field  is given by (for details see  Ref. \cite{selym1}).
\begin{eqnarray}
\begin{array}{c}\displaystyle
\mathcal{L}_{1\mathrm{R}}^{(2)}\approx-\frac{\alpha\rm m^4 \mathfrak{b}^2 }{16\pi^3}\mathcal{N}_1,\\ \\ \displaystyle
\mathcal{L}_{2\mathrm{R}}^{(2)}\approx\frac{\alpha\rm m^4 \mathfrak{b}^2}{32\pi^3}\left[\mathcal{N}_2\ln\left(\frac{\mathfrak{b}}{\gamma\pi}\right)-\frac{1}{3}\ln^2\left(\frac{\mathfrak{b}}{\gamma\pi}\right)+\mathcal{A}\right],\\ \\ \displaystyle
\mathcal{L}_{3\mathrm{R}}^{(3)}\approx\frac{\alpha\rm m^4\mathfrak{b}^2}{32\pi^3}\left[\mathcal{N}_3\ln\left(\frac{\mathfrak{b}}{\gamma\pi}\right)+\frac{1}{3}\ln^2\left(\frac{\mathfrak{b}}{\gamma\pi}\right)+\mathcal{B}\right].\end{array}\label{leading1}
\end{eqnarray} These expressions  are  accurate up to terms that decrease with $\mathfrak{b}$ like $\sim \mathfrak{b}^{-1}\ln(\mathfrak{b})$ and faster. Here  the numerical constants are
\begin{equation}
\begin{array}{c}
\mathcal{N}_{1}=1.25,\ \ \mathcal{N}_{2}=0.71, \ \ \mathcal{N}_{3}=0.29,\\
\mathcal{A}=4.21\ \ \mathrm{and}\ \ \mathcal{B}=0.69.
\end{array}
\end{equation}
Moreover, the asymptotic behavior of the full two-loop term is
\begin{equation}
\mathcal{L}_{\mathrm{R}}^{(2)}\approx\frac{\alpha\rm m^4 \mathfrak{b}^2}{32\pi^3}\left[\ln\left(\frac{\mathfrak{b}}{\gamma\pi}\right)+2.4\right],
\end{equation} which coincides with the  results reported in references
\cite{ritus,dittrich1}.

\section{Properties of a highly magnetized vacuum}

In  the presence of an  external magnetic field, the zero point vacuum
energy $ \mathscr{E}_{\mathrm{vac}}$ is  modified by the interaction
between $\textbf{B}$ and the virtual QED-particles. The latter is
determined by the effective potential coming from the
quantum-corrections to the Maxwell Lagrangian which is also
contained within the finite temperature formalism. According to Eq.
(\ref{therdp}) it is  expressed as
$\mathscr{E}_{\mathrm{vac}}=-\mathcal{L}_{\mathrm{R}}^{(1)}-\sum_{i=1}^3\mathcal{L}_{i\mathrm{R}}^{(2)}+\ldots
$  Consequently the vacuum  acquires a non trivial magnetization
$\mathscr{M}_{\mathrm{vac}}=-\partial\mathscr{E}_{\mathrm{vac}}/\partial\rm
\vert\textbf{B}\vert$ induced by the external magnetic field. In
what follows we will write
$\mathscr{M}_{\mathrm{vac}}=\mathscr{M}_{\mathrm{vac}}^{(1)}+\mathscr{M}_{\mathrm{vac}}^{(2)}+\ldots$
in correspondence with the loop-term
$\mathcal{L}^{(i)}_{\mathrm{R}}.$ In this sense, the one loop
contribution at very large magnetic field $\rm b\gg1$ can be
computed  by means of Eq. (\ref{1looprenorapp}) and gives
\begin{eqnarray}
\mathscr{M}^{(1)}=\frac{\partial\mathcal{L}_{\mathrm{R}}^{(1)}}{\partial
\vert\mathbf{B}\vert}\approx\frac{\rm m^4 \mathfrak{b}}{24\pi^2 \rm B_c}\left[2\ln\left(\frac{\mathfrak{b}}{\gamma\pi}\right)+1+\frac{12\zeta^{\prime}(2)}{\pi^2}\right].
\label{magnetization1loop}
\end{eqnarray}
Incidentally, the above  asymptotic behavior is also manifest in the corresponding magnetization derived from  the QCD vacuum in a magnetic field $\vert\mathbf{B}\vert\gg\Lambda_{\mathrm{QCD}}^2/e.$ For more details we refer the reader to  Ref. \cite{Cohen:2008bk}.

The two-loop correction is given by $\mathscr{M}^{(2)}=\sum_{i=1}^3\mathscr{M}_{i}^{(2)}$  where  $\mathscr{
M}_{i}^{(2)}=\partial\mathcal{L}_{i\mathrm{R}}^{(2)}/\partial\vert\textbf{B}\vert$
is  the contribution corresponding to a  photon propagation
mode. Making use of Eqs. (\ref{leading1}) we find
\begin{eqnarray}
\mathscr{M}_1^{(2)}&\approx&-\frac{\alpha\rm m^4\mathfrak{b}}{8 \pi^3 \rm B_c}\mathcal{N}_1,\\ \displaystyle
\mathscr{M}_2^{(2)}&\approx&-\frac{\alpha\rm m^4\mathfrak{b}}{32 \pi^3 \rm B_c}\left[\frac{2}{3}\ln^2\left(\frac{\mathfrak{b}}{\gamma \pi}\right)\right.\nonumber\\&&+\left.\frac{8\zeta^{\prime}(2)}{\pi^2}\ln\left(\frac{\mathfrak{b}}{\gamma \pi}\right)-9.13\right],\\ \displaystyle
\mathscr{M}_3^{(2)}&\approx&\frac{\alpha\rm m^4\mathfrak{b}}{32 \pi^3 \rm B_c}\left[\frac{2}{3}\ln^2\left(\frac{\mathfrak{b}}{\gamma \pi}\right)\right.\nonumber\\&&+\left.\left(2+\frac{8\zeta^{\prime}(2)}{\pi^2}\right)\ln\left(\frac{\mathfrak{b}}{\gamma \pi}\right)+1.67\right].
\label{leadingmag1}
\end{eqnarray}

According to these results,  in a superstrong magnetic field approximation,
$\mathscr{M}_1^{(2)}<0$ and $\mathscr{M}_2^{(2)}<0$ behave diamagnetically
whereas  $\mathscr{M}_3^{(2)}>0$ is purely  paramagnetic. Moreover, while   $\mathscr{M}_1^{(2)}$  depends linearly on $\mathfrak{b},$ the  contributions of the second and third propagation mode depend
logarithmically on   the external field. We find, in particular, that for magnetic fields  $\rm
B\sim10^{18}G,$ the magnetization  generated by  the first and
second polarization mode  reaches values  the order
$\sim-10^{12}\rm erg/(cm^3 G)$ and  $\sim -10^{13}\rm erg/(cm^3G),$
respectively.  In the same context
$\mathscr{M}^{(3)}\sim+10^{13}\rm erg/(cm^3G).$ Note that the leading
behavior of the complete two-loop contribution is
\begin{equation}
\mathscr{M}^{(2)}\approx\frac{\alpha\rm m^4 \mathfrak{b}}{16\pi^3\rm B_c}\left[\ln\left(\frac{\mathfrak{b}}{\gamma\pi}\right)+2.9\right]>0.
\end{equation} which shows a  dominance of the third mode. Indeed, for  $\rm
B\sim10^{18}G,$  one finds  $\mathscr{M}^{(2)}\sim+10^{12}\rm erg/(cm^3G).$

As it was expected  $\mathscr{M}^{(1)}/\mathscr{M}^{(2)}\sim\alpha^{-1}.$ This ratio
is  also manifested between the  corresponding  magnetic susceptibilities ($\mathscr{X}^{(i)}=\partial
\mathscr{M}^{(i)}/\partial \vert\mathbf{B}\vert$). Note that
\begin{eqnarray}
\mathscr{X}^{(1)}&\approx&\frac{\rm m^4}{24 \pi^2 \rm
\rm B_c^2}\left[2\ln\left(\frac{\mathfrak{b}}{\gamma\pi}\right)+1.86\right]>0,\\
\mathscr{X}^{(2)}&\approx&\frac{\alpha \rm m^4 }{16 \pi^3 \rm \rm
B_c^2}\left[\ln\left(\frac{\mathfrak{b}}{\gamma\pi}\right)+3.9\right]>0.
\end{eqnarray}
For magnetic fields $\mathfrak{b}\sim 10^5$ corresponding to  $\rm \vert\textbf{B}\vert\sim 10^{18}G,$  the
magnetic susceptibility reaches   values of the order of
$\mathscr{X}^{(1)}\sim10^{-4}\rm  erg/(cm^3 G^2)$ which exceeds the
values of many laboratory materials, for example   Aluminum ($\mathscr{X}_{\mathrm{Al}}=2.2\cdot 10^{-5}\rm erg/(cm^3 G^2)$).

Some additional comments are in order. First of all, even though the previous decomposition of  $\mathscr{M}^{(2)}$ is  not really observable, it turns out to be a transparent framework which  illustrates, in a  phenomenological way,  the  magnetic property generated by each photon propagation mode and thus, a connection with the birefringent property of the vacuum. The decomposition of $\mathscr{M}^{(2)}$ in terms of the photon modes allows, in addition, to establish similarities and differences with the magnetization carried by observable photons\footnote{The magnetic response of an observable photon was studied in \cite{hugoe} in  two different regimes of the vacuum polarization tensor.  On the one
hand for low  energies in weak fields ($\vert\bf B\vert\ll\rm B_c$)
and on the other hand (originally studied in Ref. \cite{selym})
near the first pair creation threshold and for a moderate fields
$(\vert\bf B\vert\sim \rm B_c)$.}.  Indeed, similar to the latter  the virtual radiation carries a  magnetization which depends on the polarization vector. However, while an observable second  mode  has a   paramagnetic response \cite{hugoe}, our result  points  out that  the   corresponding virtual  polarization   generates a purely diamagnetic magnetization. Besides, we have seen  that the first propagation mode contributes to the magnetization of the vacuum. This is not expected for  an observable mode-1 photon since its dispersion law is independent on the external field strength \cite{shabad2,shabad6,Shabad:2009ky} and therefore does not carry a  magnetization.

Because of the anisotropy generated by $\bf B$ a magnetized vacuum
exerts two different pressures
\cite{Chaichian:1999gd,Martinez:2003dz,PerezRojas:2004ip,PerezRojas:2004in}.
One of them is positive
$(\mathscr{P}_\parallel=-\mathscr{E}_{\mathrm{vac}})$ and along
$\textbf{B},$ whereas the remaining is  transverse to the external
field direction
($\mathscr{P}_\perp=-\mathscr{E}_{\mathrm{vac}}-\mathscr{M}\vert\textbf{B}\vert$).
For $\mathfrak{b}\sim 1$  the  latter  acquires  negative values. At very
large magnetic  fields ($\mathfrak{b}\gg1$) the one-loop approximation of
$\mathscr{P}_\perp$ can be computed by making use of Eq.
(\ref{1looprenorapp}) and Eq. (\ref{magnetization1loop}).  In fact
\begin{equation} \mathscr{P}_\perp^{(1)}\approx-\frac{\rm m^4\rm \mathfrak{b}^2}{24
\pi^2}\left[\ln\left(\frac{\mathfrak{b}}{\gamma
\pi}\right)+1+\frac{6\zeta^\prime(2)}{\pi^2}\right]<0.
\end{equation} Therefore,  at  asymptotically  large  values of the external field,
the interaction between $\bf B$ and the virtual  electron positron pairs  generates a negative pressure which would tend to shrink inserted matter in the plane transverse  to  $\textbf{B}.$

Again, the two-loop contribution  can be written as the sum of the
corresponding  terms due to the  vacuum polarization modes
$\mathscr{P}_{\perp}^{(2)}=\sum_{i=1}^3\mathscr{P}_{\perp i}^{(2)}. $ According to Eqs. (\ref{leading1}) and Eqs.
(\ref{leadingmag1}) they read:
\begin{eqnarray}
\mathscr{P}_{\perp1}^{(2)}&\approx&\frac{\alpha\rm m^4\mathfrak{b}^2}{16 \pi^3}\mathcal{N}_1>0, \\    \mathscr{P}_{\perp2}^{(2)}&\approx&\frac{\alpha \rm m^4\rm \mathfrak{b}^2}{32 \pi^3}\left[\frac{1}{3}\ln^2\left(\frac{\mathfrak{b}}{\gamma\pi}\right)+\left(\mathcal{N}_2+\frac{8\zeta^{\prime}(2)}{\pi^2}\right)\right.\nonumber\\ &&\times \left.\ln\left(\frac{\mathfrak{b}}{\gamma\pi}\right)-4.92\right]>0, \\
\mathscr{P}_{\perp3}^{(2)}&\approx&-\frac{\alpha \rm m^4\mathfrak{b}^2}{32 \pi^3}\left[\frac{1}{3}\ln^2\left(\frac{\mathfrak{b}}{\gamma\pi}\right)+\left(1.71+\frac{8\zeta^{\prime}(2)}{\pi^2}\right)\right.\nonumber\\ &&\times\left.\ln\left(\frac{\mathfrak{b}}{\gamma\pi}\right)+0.98\right]<0,
\end{eqnarray}
with the complete two-loop term given by
\begin{equation}
\mathscr{P}_{\perp}^{(2)}\approx -\frac{\alpha\rm m^4
\mathfrak{b}^2}{32\pi^3}\left[\ln\left(\frac{\mathfrak{b}}{\gamma\pi}\right)+3.4\right]<0.
\end{equation}
For $\mathfrak{b}\sim 10 ^5,$ corresponding to magnetic fields  $\rm
B\sim10^{18}G,$ the transverse pressure generated by  the first and
second polarization mode are positive and reaches values of the order
$\sim10^{30}\rm dyn/cm^2$ and  $\sim 10^{31}\rm dyn/cm^2,$
respectively. In contrast, the contribution given by the  third
mode is negative with   $\mathscr{P}_{\perp3}^{(2)}\sim-10^{31}\rm
dyn/cm^2.$  The combined result  is negative and achieves  values of the order  $\mathscr{P}_{\perp}^{(2)}\sim-10^{31}\rm dyn/cm^2.$ This  fact strengthens  the analogy of the considered problem  with the  Casimir effect in which  the  pressure transversal  to the parallel plates is also negative and  dominated by  the virtual electromagnetic radiation \cite{milonni}.

\section{The vacuum instability in a supercritical\\ electric field}

Ultra-high electric  fields  $\vert\mathbf{E}\vert\gg \mathrm{E}_c=\mathrm{m}^2/\mathrm{e}=1.3\cdot10^{16}\mathrm{V}/\mathrm{cm}$ have been predicted  to exist at the surface of strange stars \cite{AFO86,AFO861,usov-ef,usov-ef1}. In this electric field regime,  the asymptotic behavior of $\mathcal{L}_{i\mathrm{R}}^{(2)}$  is obtained from Eqs. (\ref{leading1}) by means of the duality  transformation $\mathfrak{b}\to -i\mathfrak{e}$ with $\mathfrak{e}=\vert\mathbf{E}\vert/\mathrm{E}_c.$ As a consequence we  can write
\begin{equation}
\mathcal{L}_{\mathrm{EH}}=\Re\left[\mathcal{L}_{\mathrm{EH}}\right]+\mathrm{Im}\left[\mathcal{L}_{\mathrm{EH}}\right]\label{imaginariaeuhei}
\end{equation}where $\Re[\mathcal{L}_{\mathrm{EH}}]$ constributes to the dispersive effects.
Because of the imaginary part the vacuum becomes unstable and creation of particles could take place. The probability associated with the vacuum decay is $
\mathcal{P}=1-\vert\langle0_\mathrm{out}\vert0_\mathrm{in}\rangle\vert^2$  with $\langle0_\mathrm{out}\vert0_\mathrm{in}\rangle=\mathrm{e}^{i\mathrm{VT}\mathcal{L}_{\mathrm{EH}}},$ where $\rm VT$ is the  volume element in $3+1$ dimensions. With this in mind and by considering Eq. (\ref{imaginariaeuhei}) one has
\begin{equation}
\mathcal{P}=1-\mathrm{e}^{-2\mathrm{Im}\left[\mathcal{L}_{\mathrm{EH}}\right]\mathrm{VT}}.\label{rate43}
\end{equation} As we have already mentioned in the introduction,   the emission of observable photons from the vacuum is also a plausible effect. The probability $\mathfrak{P}$ of photon emission from the vacuum  accompanied by the creation of an arbitrary number of electron-positron pairs, can be  determined by using  the unitarity condition for the  dispersion matrix $\mathcal{S}=1+i\mathscr{T}$ which leads to a relation of  the optical-theorem type \cite{Gitman:1986xr,gitman1,gitman2}
\begin{equation}
\sum_{\mathrm{out}}\left\vert\langle\mathrm{out}\vert\mathscr{T}\vert\mathrm{in}\rangle\right\vert^2=2\mathrm{Im}\langle\mathrm{in}\vert\mathscr{T}\vert\mathrm{in}\rangle,\label{optical}
\end{equation} where $\mathscr{T}$ is the sum of all  Feynman graphs. However, the electron causal Green function for the matrix elements  $\langle\mathrm{in}\vert\ldots\vert\mathrm{in}\rangle$  is quite different from the standard propagator in the Schwinger proper-time representation of the out-in matrix elements.  We consider  the relation above  with the $\mathrm{in}-$state having no photons \emph{i.e.} $\vert\mathrm{in}\rangle=\vert0_{\mathrm{in}}\rangle$. In addition, we will confine ourselves in the right- and left-hand sides of Eq. (\ref{optical}) to the second order radiational interaction.  In this approximation, one can  take the operator $\mathscr{T}$ on the left-hand side to first order: $\mathscr{T}^{(1)}=-\int d^4x j^\mu(x)a_\mu(x).$ Here  $j^\mu=\frac{\mathrm{e}}{2}\left[\bar{\psi}(x)\gamma^\mu,\psi(x)\right]$ is the current and  $a_\mu(x)$ denotes  the radiation field. Therefore, only one-photon states contribute to the sum over the $\mathrm{out}-$states
$\langle\mathrm{out}\vert=\langle0_{\mathrm{in}}\vert c_{\boldsymbol{k}i}\-b_{\boldsymbol{q}_1}\ldots b_{\boldsymbol{q}_n}d_{\boldsymbol{p}_1}\ldots d_{\boldsymbol{p}_n}.$ Whilst $c_{\boldsymbol{k}i}$ is understood  as the annihilation operator of a mode-$i$ photon,  $b$ and $d$ are interpreted as the annihilation operators of electrons and positrons, respectively.  On the right-hand side of Eq. (\ref{optical}) $\mathscr{T}$ has to be taken in second order $\mathscr{T}^{(2)}=-\frac{i}{2}\int d^4x \-d^4x^\prime T\left[j^\mu(x)j^\nu(x^\prime)a_\mu(x)a_\nu(x^\prime)\right],$ where $T$ represents the time ordering operator. Considering the normal mode expansions of the dynamical fields  and by using the Wick theorem one obtains
\begin{equation}
\langle\mathrm{in}\vert\mathscr{T}^{(2)}\vert\mathrm{in}\rangle=\mathcal{L}_{\mathrm{in}}^{(2)}
\end{equation} where  $\mathcal{L}_{\mathrm{in}}^{(2)}$ is determined by  Eq. (\ref{twoloopunrenor}) with the causal Feynman propagators replaced by   the corresponding electron Green function appearing in the mean values $\langle\mathrm{in}\vert\ldots\vert\mathrm{in}\rangle$ of Eq. (\ref{optical}).

Now the  decay probability $\mathcal{P}$ is connected to the total probability of photon emission from the vacuum, accompanied by the creation of an arbitrary number of electron-positron pairs  $\mathcal{P}=\mathfrak{P}+\ldots$ \footnote{For details, see the cited references of  Fradkin, Gitman, Shvartsman.} Obviously, the corresponding expansion upto second order in $\alpha$ involves the imaginary part coming from the two-loop contribution of the Euler-Heisenberg Lagrangian. The latter can be written as the sum of the  corresponding  terms due to the individual  vacuum polarization modes $\mathrm{Im}\left[\mathcal{L}_{\mathrm{R}}^{(2)}\right]=\sum_{i=1}^3\mathrm{Im}\left[\mathcal{L}_{i\mathrm{R}}^{(2)}\right]$ with
\begin{eqnarray}\label{imehl}
\mathrm{Im}\left[\mathcal{L}_{1\mathrm{R}}^{(2)}\right]&\approx&0,\nonumber\\  \mathrm{Im}\left[\mathcal{L}_{2\mathrm{R}}^{(2)}\right]&\approx&\frac{\alpha\rm m^4
\mathfrak{e}^2}{32\pi^2}\left\{\frac{1}{2}\mathcal{N}_2-\frac{1}{3}\ln\left(\frac{\mathfrak{e}}{\gamma\pi}\right)\right\}<0\\ \mathrm{Im}\left[\mathcal{L}_{3\mathrm{R}}^{(2)}\right]&\approx&\frac{\alpha\rm m^4
\mathfrak{e}^2}{32\pi^2}\left\{\frac{1}{2}\mathcal{N}_3+\frac{1}{3}\ln\left(\frac{\mathfrak{e}}{\gamma\pi}\right)\right\}>0. \nonumber
\end{eqnarray} Note that,  because of the  dominance of the  third propagation mode,  the complete imaginary part arising from the  two-loop term  is positive
\begin{equation}
\mathrm{Im}\left[\mathcal{L}^{(2)}_{\mathrm{R}}\right]\approx\frac{3\alpha}{4\pi}\mathrm{Im}\left[\mathcal{L}_{\mathrm{R}}^{(1)}\right]>0
\end{equation} where  $\mathrm{Im}\left[\mathcal{L}_{\mathrm{R}}^{(1)}\right]\simeq \frac{\rm m^4 \mathfrak{e}^2}{48 \pi}$ is the imaginary part coming from the  one-loop contribution at very large electric field $\mathfrak{e}\gg1.$

There is some interesting aspects in Eq. (\ref{imehl})  which deserves some comments. First of all, it reveals that the  first propagation mode  does not contribute to the imaginary part of  $\mathcal{L}_{\mathrm{EH}}$. This suppression provides an evidence that the vacuum does not generate electromagnetic waves  propagating along the external field $\bf E$. On the other hand, only the second and third mode contribute to the vacuum instability. This is a signal that the vacuum could  create the corresponding  propagation modes. However, the birefringence character  of the vacuum  imposes that there is not a single observable mode-2 photon without the existence  of a corresponding mode-3 photon. Therefore, in addition to the electron-positron pairs, the vacuum could create  photons with  different propagation modes. Note, however,  that the described photon emission  has been  predicted   within the framework of equilibrium  quantum field theory, even though it is a far-from-equilibrium, time-dependent phenomenon. Only further studies can tell us how far this process can be stretched because a realistic treatment of this issue  requires   a time evolution analysis of the photon number distribution functions, similar to that developed by Hebenstreit  \emph{et al.} for  electron-positron pairs within a quantum kinetic approach \cite{Hebenstreit:2008ae}.

\section{Summary and Outlook}

In summary, we have examined the magnetization of the QED vacuum in the presence of a constant magnetic field in the strong field regime, $\vert\mathbf{B}\vert\gg \rm B_c.$ We have seen  that the virtual electromagnetic radiation is a
source of magnetization to the whole vacuum. In a superstrong  magnetic field approximation,
the two-loop contribution of the magnetization density corresponding to
the  second and third propagation mode depends  nonlinearly on the external magnetic field and their
behavior is diamagnetic and paramagnetic, respectively. On the other hand,
the contribution coming from the first mode is
diamagnetic and depends linear on  $\rm B$.  We have seen
that for very large magnetic field the contribution of  the
third mode strongly dominates the analyzed quantities. In this  regime the latter tends  to  shrink  inserted matter
by exerting a negative transverse pressures to the external field. On
the contrary  those contributions coming from the first and second
virtual mode are  positive and tend to expand the matter.

In the  last section of this Letter we showed that  only two photon
propagation modes contribute to the instability of the vacuum in an
strong electric field. This  instability is  associated  with the
emission of  photon whose  propagation modes differ each other. A
plausible connection between this mechanism and the birefringent
character of the vacuum occupied by a supercritical electric field
was discussed and the suppression of vacuum decay into pair of modes
propagating along an external electric field was analyzed as well.
\vspace{0.005 in}
\begin{flushleft}
\textbf{Acknowledgments}
\end{flushleft}
\vspace{0.005 in} The author would like  to  expresses its deep
gratitude to professor A.  Shabad for  valuable suggestions and
advice. He also thanks F. Hebenstreit, Helios Sanchis Alepuz  and
Kai Schwenzer for several discussions and important remarks.  The
author wants  to extend his gratitude to professor Reinhard Alkofer
for helping to  support this research. This work  has been supported
by the Doktoratskolleg ``Ha\-drons in Vacuum, Nuclei and Stars''  of
the Austrian Science Fund (FWF) under contract W1203-N08.

\end{document}